\documentstyle[12pt,epsfig]{article}

\catcode`\@=11
\long\def\@makefntext#1{
\protect\noindent \hbox to 3.2pt {\hskip-.9pt  
$^{{\ninerm\@thefnmark}}$\hfil}#1\hfill}		

\def\@makefnmark{\hbox to 0pt{$^{\@thefnmark}$\hss}}  
	
\def\ps@myheadings{\let\@mkboth\@gobbletwo
\def\@oddhead{\hbox{}
\rightmark\hfil\ninerm\thepage}   
\def\@oddfoot{}\def\@evenhead{\ninerm\thepage\hfil
\leftmark\hbox{}}\def\@evenfoot{}
\def\sectionmark##1{}\def\subsectionmark##1{}}

\renewcommand{\thefootnote}{\fnsymbol{footnote}}

\newcounter{sectionc}\newcounter{subsectionc}\newcounter{subsubsectionc}
\renewcommand{\section}[1] {\vspace*{0.6cm}\addtocounter{sectionc}{1} 
\setcounter{subsectionc}{0}\setcounter{subsubsectionc}{0}\noindent 
	{\normalsize\bf\thesectionc. #1}\par\vspace*{0.4cm}}
\renewcommand{\subsection}[1] {\vspace*{0.6cm}\addtocounter{subsectionc}{1} 
	\setcounter{subsubsectionc}{0}\noindent 
	{\normalsize\it\thesectionc.\thesubsectionc. #1}\par\vspace*{0.4cm}}
\renewcommand{\subsubsection}[1]
{\vspace*{0.6cm}\addtocounter{subsubsectionc}{1}
	\noindent {\normalsize\rm\thesectionc.\thesubsectionc.\thesubsubsectionc. 
	#1}\par\vspace*{0.4cm}}

\newcounter{appendixc}
\newcounter{subappendixc}[appendixc]
\newcounter{subsubappendixc}[subappendixc]

\renewcommand{\appendix}[1] {\vspace*{0.6cm}
        \refstepcounter{appendixc}
        \setcounter{figure}{0}
        \setcounter{table}{0}
        \setcounter{equation}{0}
        \renewcommand{\thefigure}{\Alph{appendixc}.\arabic{figure}}
        \renewcommand{\thetable}{\Alph{appendixc}.\arabic{table}}
        \renewcommand{\theappendixc}{\Alph{appendixc}}
        \renewcommand{\theequation}{\Alph{appendixc}.\arabic{equation}}
        \noindent{\bf Appendix \theappendixc #1}\par\vspace*{0.4cm}}

\def\abstracts#1{{
	\centering{\begin{minipage}{12.2truecm}\footnotesize\baselineskip=12pt\noindent
	\centerline{\footnotesize ABSTRACT}\vspace*{0.3cm}
	\parindent=0pt #1
	\end{minipage}}\par}} 

\renewenvironment{thebibliography}[1]
	{\begin{list}{\arabic{enumi}.}
	{\usecounter{enumi}\setlength{\parsep}{0pt}
\setlength{\leftmargin 1.25cm}{\rightmargin 0pt}
	 \setlength{\itemsep}{0pt} \settowidth
	{\labelwidth}{#1.}\sloppy}}{\end{list}}

\newcounter{itemlistc}
\newcounter{romanlistc}
\newcounter{alphlistc}
\newcounter{arabiclistc}

\newcommand{\fcaption}[1]{
        \refstepcounter{figure}
        \setbox\@tempboxa = \hbox{\footnotesize Fig.~\thefigure. #1}
        \ifdim \wd\@tempboxa > 6in
           {\begin{center}
        \parbox{6in}{\footnotesize\baselineskip=12pt Fig.~\thefigure. #1}
            \end{center}}
        \else
             {\begin{center}
             {\footnotesize Fig.~\thefigure. #1}
              \end{center}}
        \fi}

\newcommand{\tcaption}[1]{
        \refstepcounter{table}
        \setbox\@tempboxa = \hbox{\footnotesize Table~\thetable. #1}
        \ifdim \wd\@tempboxa > 6in
           {\begin{center}
        \parbox{6in}{\footnotesize\baselineskip=12pt Table~\thetable. #1}
            \end{center}}
        \else
             {\begin{center}
             {\footnotesize Table~\thetable. #1}
              \end{center}}
        \fi}

\def\@citex[#1]#2{\if@filesw\immediate\write\@auxout
	{\string\citation{#2}}\fi
\def\@citea{}\@cite{\@for\@citeb:=#2\do
	{\@citea\def\@citea{,}\@ifundefined
	{b@\@citeb}{{\bf ?}\@warning
	{Citation `\@citeb' on page \thepage \space undefined}}
	{\csname b@\@citeb\endcsname}}}{#1}}

\newif\if@cghi
\def\cite{\@cghitrue\@ifnextchar [{\@tempswatrue
	\@citex}{\@tempswafalse\@citex[]}}
\def\citelow{\@cghifalse\@ifnextchar [{\@tempswatrue
	\@citex}{\@tempswafalse\@citex[]}}
\def\@cite#1#2{{$\null^{#1}$\if@tempswa\typeout
	{IJCGA warning: optional citation argument 
	ignored: `#2'} \fi}}

 1
 1
 1

\font\ninerm=cmr9

\def\frac#1#2{ {{#1} \over {#2} }}

\def\abs#1{\left| \: #1 \: \right|}%
\def\bom#1{\mbox{\boldmath$#1$}}
\def\beq{\begin{displaymath}}
\def\eeq{\end{displaymath}}

\def\as{\alpha_S}
\def\asb{\bar \alpha_S}

\def\bkq{\abs{\bom{k}+\bom{q}}}
\def\om{\omega}
\def\ga{\gamma}
\def\tga{\tilde \gamma}
\def\tchi{\tilde \chi}

\def\De{\Delta}

\def\cA{{\cal A}}
\def\cP{{\cal P}}
\def\Q_s{\mu}

\def\np#1#2#3{Nucl.\ Phys.\ B#1 (19#3) #2}
\def\pl#1#2#3{Phys.\ Lett.\ #1B (19#3) #2}
\def\pr#1#2#3{Phys.\ Rev.\ D #1 (19#3) #2}

\def\zp#1#2#3{Zeit.\ Phys.\ C#1 (19#3) #2}

\begin{document}
\begin{flushright}
IFUM-577-FT \\
hep-ph/9707383
\vspace{0.5cm}
\end{flushright}
\centerline{\normalsize\bf ANGULAR ORDERING AND SMALL-$\bom{x}$ 
EVOLUTION\footnote{Talk presented at the Ringberg workshop: New trends
in HERA physics, Germany, May 1997} 
}
\baselineskip=16pt
\centerline{\footnotesize GAVIN P. SALAM}
\baselineskip=13pt
\centerline{\footnotesize\it INFN --- Sezione di Milano, Via Celoria
16}
\baselineskip=12pt
\centerline{\footnotesize\it  20133 Milano, Italy}
\centerline{\footnotesize E-mail: Gavin.Salam@mi.infn.it}
\vspace*{0.3cm}

\vspace*{0.9cm}
\abstracts{This talks examines the effect of angular ordering on the
small-$x$ evolution of the unintegrated gluon distribution, and
discusses the characteristic function for the CCFM equation, as well as
some preliminary results on final-state properties.}
 
\normalsize\baselineskip=15pt
\setcounter{footnote}{0}
\renewcommand{\thefootnote}{\alph{footnote}}
\section{Introduction}
For some time now it has been known that angular
ordering\cite{yuribook} is an essential element in any description of
small-$x$ final state properties.\cite{CCFMa,CCFMb} As a first step of a
programme to study the final state in small-$x$ physics, one should
examine the effect of angular ordering on the small-$x$ evolution of
the gluon structure function. Phenomenological studies have already
been performed,\cite{Durham} but this talk will examine the solutions
of the CCFM equation\cite{CCFMa,CCFMb} from a more theoretical point of
view. I will also present some first results on final-state properties.

The main difference between the BFKL\cite{BFKL} and CCFM equations is
in the collinear region: in the BFKL case, the $i^{th}$ emission has a
transverse momentum $q_i >\mu$, with $\mu$ a cutoff put in by hand and
which is taken to zero; this regulates the collinear divergence, which
for the gluon structure function cancels out. But in quantities where
it doesn't cancel, such as certain final properties, one obtains the
wrong answer. In the CCFM equation, angular ordering of emissions
leads to the following condition (see figure~\ref{figkinebw}):

\begin{displaymath}\label{ao}
\theta_{i}>\theta_{i-1}\,,
\;\;\;\;\;\;\Rightarrow\;\;\;\;\;\;
q_{i} > z_{i-1} q_{i-1}
\,,
\end{displaymath}

\noindent with the corresponding gluon emission distribution being

\begin{displaymath}\label{dP}
d{\cal P}_i
=\frac{d^2q_i}{\pi q_i^2} \; dz_i\frac{\asb}{z_i}
\;\De(z_i,q_i,k_i)\;\Theta(q_i-z_{i-1}q_{i-1})
\,.
\end{displaymath}

\begin{figure}[t]
\hspace{2.6cm}\input{kine_bw.pstex_t}
\fcaption{Labelling of momenta.}
\label{figkinebw}
\end{figure}

\noindent The non-Sudakov form factor $\De$, which is analogous to a
probability for suppressing any further radiation, is defined by 

\begin{displaymath}\label{De}
\ln \De(z_i,q_i,k_i)=
-\int_{z_i}^1 dz' \;\frac{\asb}{z'}
\int\frac{dq'^2}{q'^2}\;\Theta(k_i-q')\;\Theta(q'-z'q_i)
\,.
\end{displaymath}

The elimination of a large fraction of the small-transverse-momentum
emissions means that angular ordering has a big effect on the final
state. But in structure function evolution, since collinear
singularities cancel, at leading order the BFKL and CCFM structure
functions are equivalent.

\section{The gluon structure function}
As part of a program to carry out a full investigation of the effects
of angular ordering at small $x$, this talk examines the component of
the next-to-leading order corrections to structure function evolution
that arise from angular ordering. Such effects are expected to be part
of the full NLL contribution.\cite{nllsmlx}

Qualitatively, since angular ordering reduces the phase space for
evolution the exponent of the small-$x$ growth ought to be
reduced. The symmetry, present in the BFKL equation, between large and
small scales will be broken, favouring evolution to large momentum
scales. Finally diffusion will be reduced because large jumps (down)
in scale are suppressed.

There are two limits in which, at small $x$, the effects of angular
ordering should disappear: as $\as \to 0$, because the typical
$z_{i-1} \sim \as$ will be very small (this justifies the assertion
that for structure functions the effects of angular ordering are
next-to-leading); and in the double-leading-logarithmic limit because
the condition $q_i > q_{i-1}$ automatically satisfies the angular
ordering condition.

The analytic treatment of the CCFM equation is more complicated than
that of the BFKL equation because the gluon density contains one
extra parameter, $p$, which defines the maximum angle for the emitted
gluons. In DIS it enters through the angle of the quarks produced
in the boson-gluon fusion. The equation for the CCFM density,
$A(x,k,p)$, of gluons with longitudinal momentum fraction $x$ and
transverse momentum $k$ is:

\begin{displaymath}\label{A1}
\cA(x,k,p) \;=\; \cA^{(0)}(x,k,p) \;+\;
\int\frac{d^2q}{\pi q^2}\; \frac{dz}{z}
\;\frac{\asb}{z}\De(z,q,k)\;
\Theta(p-zq)\;\cA(x/z,k',q)
\,,
\end{displaymath}

\noindent where $k'=\bkq$. By analogy to the BFKL equation one can
develop some understanding of it by looking for eigensolutions
(strictly speaking eigensolutions of the equation without an
inhomogeneous term and with no lower limit in the $z$ integral) of the
form

\begin{displaymath}
x\cA(x,k,p) = 
x^{-\om} \frac{1}{k^2}\left(\frac{k^2}{k_0^2}\right)^{\tga} G(p/k)
\,,
\end{displaymath}

\noindent where $G(p/k)$ parameterises the unknown dependence on
$p$. For $0<\tga<1$, one obtains a coupled pair of equations for 
$G$ and $\om$:

\begin{displaymath}\label{dG}
p\;\partial_p\;G(p/k)=
\asb
\int_p\;\frac{d^2q} {\pi q^2}
\left(\frac{p}{q}\right)^{\om}\;\De(p/q,q,k)
\;G\left(\frac{q}{k'}\right)
\;\left(\frac{{k'}^2}{k^2}\right)^{\tga-1}
,
\end{displaymath}

\noindent with the initial condition $G(\infty)=1$ and

\begin{displaymath}\label{A5}
\om = \as\tchi(\tga,\as) = \as
\int\frac{d^2q}{\pi q^2}
\left\{
\left(\frac{{k'}^2}{k^2}\right)^{\tga-1}
  G\left(\frac{q}{k'}\right)
-\Theta(k-q)\; G(q/k)\right\}
.
\end{displaymath}

\begin{figure}
\begin{center}
\epsfig{file=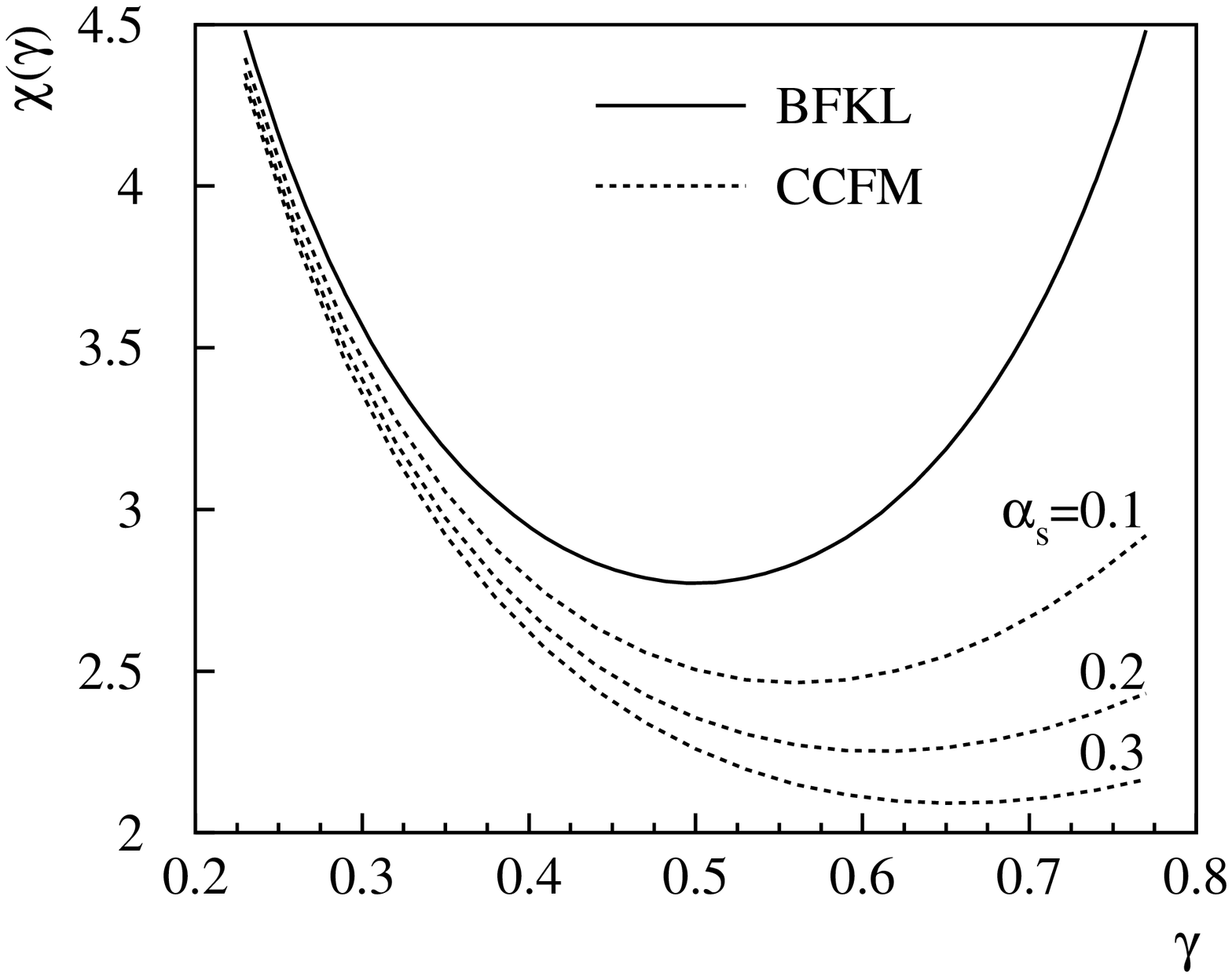, width=0.7\textwidth}
\end{center}
\fcaption{The BFKL and CCFM characteristic functions as a function of
$\ga$ for different values of $\as$.}
\label{figcvg}
\end{figure}

\noindent In the second of these equations, if $G=1$
one notes that $\tchi$ is just the BFKL characteristic function,
$\chi$. Since $1-G(p/k)$ is formally of order $\as$, this demonstrates
that angular ordering has a next-to-leading effect on structure
function evolution. One can also show that in the limit of $\ga \to 0$
the difference $\chi(\ga) - \tchi(\ga,\as)$ tends to a constant, which
implies corrections to the small-$x$ anomalous dimension of the form
$\as^3/\om^2$.

\begin{figure}
\begin{center}
\epsfig{file=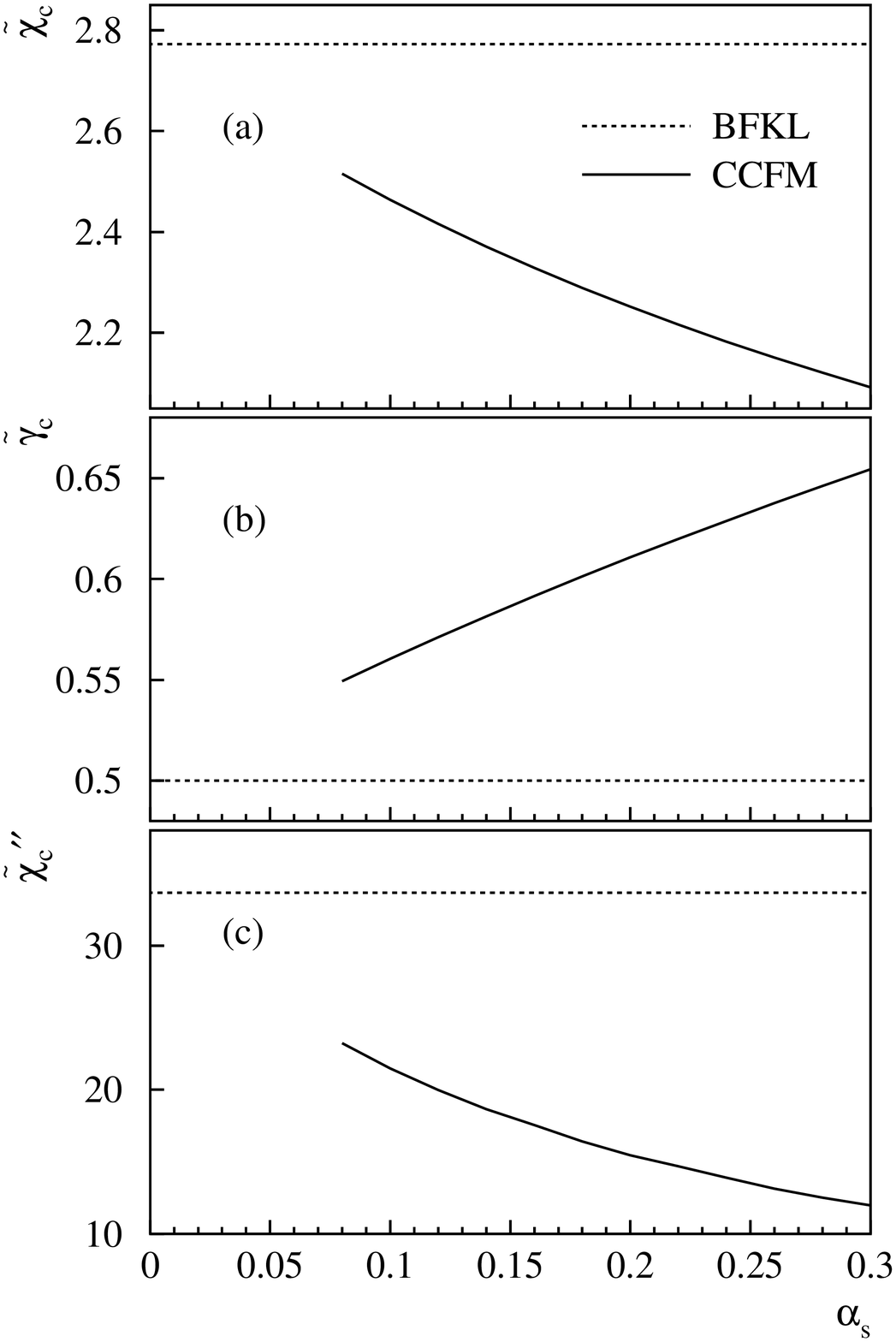, width=0.7\textwidth}
\end{center}
\fcaption{(a) The value of the minimum of the characteristic function,
 $\tchi_c$, as a function of $\as$; (b) The position of the minimum of
 the characteristic function, $\tga_c$, as a function of $\as$;
 (c) The second derivative of the characteristic function,
 ${\tchi_c}''$, at its minimum, as a function of $\as$.}
\label{fig:minva}
\end{figure}

Though a number of asymptotic properties of $G(p/k)$ have been
determined,\cite{CCFMa,bmss} it has not so far been possible to obtain
its full analytic form. Further understanding requires numerical
analysis. This has been carried out and figure~\ref{figcvg} shows
the results for $\tchi$ compared to the BFKL characteristic function
for three different values of $\as$. It illustrates that as $\as\to0$
the two tend to coincide as happens also in the region $\ga \to0$ (the
DLLA region).

The loss of symmetry under $\ga\to1-\ga$ relates to the loss of
symmetry between small and large scales. Indeed, in contrast to the
BFKL case, there is no longer even a divergence at $\ga=1$.
Correspondingly, the minimum of the characteristic function gets
shifted to the right and is lower.

Figure~\ref{fig:minva} shows the characteristics of the minimum (or
critical point) of $\tchi$ as a function of $\as$: the height of the
minimum, $\tchi_c$, its position, $\tga_c$ and the second derivative
of $\tchi$ at the minimum, $\tchi_c''$. One notes that the dependence
of these quantities on $\as$ is noticeably non-linear, indicating the
presence of substantial corrections beyond next-to-leading
logarithms. Indeed, for $\as=0.2$, carrying out a simple fit to the
linear component of the correction at small $\as$, one sees that
contributions beyond NLL are of the order of about half the total
correction. 

\begin{figure}
\begin{center}
\epsfig{file=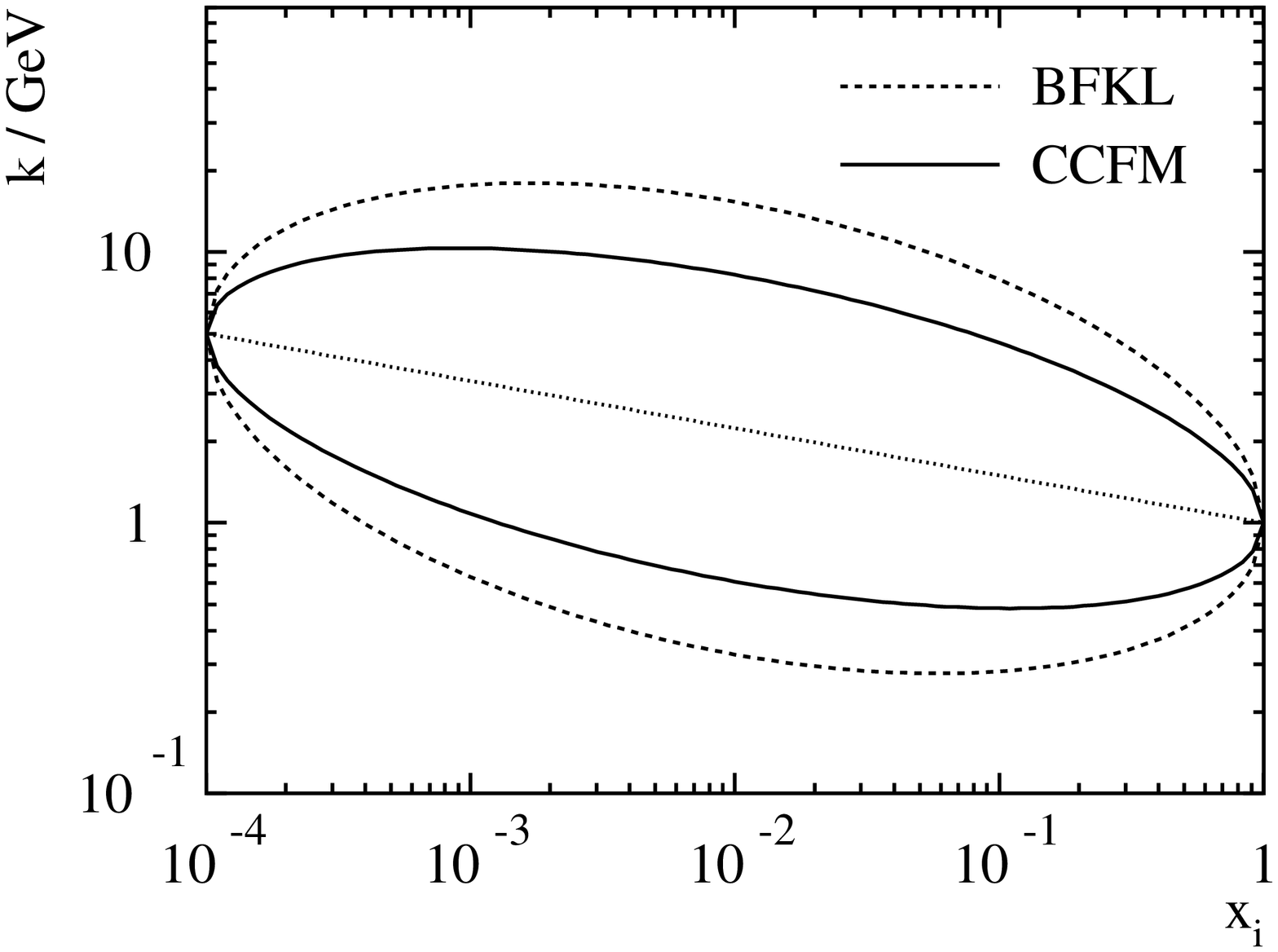,width=0.7\textwidth}
\end{center}
\fcaption{``Cigars'' showing the range of transverse momentum
in the evolution as a function of intermediate $x_i$ for the BFKL and
CCFM equations. The evolution is to $x=10^{-4}$, $k=5$~GeV with
$\as=0.2$.}
\label{fig:dfsn}
\end{figure}

Angular ordering has a particularly large effect on the second
derivative of $\tchi$ --- for $\as=0.2$, it is reduced by a factor of
two. The importance of the second derivative is that it is related to
the amount of diffusion that is present. For an evolution over a range
$x$, if one examines intermediate gluons with longitudinal momentum
fraction $x_i$, using the saddle-point approximation one
finds that the distribution of their transverse momentum has a width
$\Delta \ln k$ which is given by:

\begin{equation}
    \De \ln k \simeq \sqrt{ \frac{\asb {\chi_c}''}{4} \;
                     \frac{\ln x_i \ln x_i/x}{\ln x}}
\end{equation}

\noindent Figure~\ref{fig:dfsn} shows the application of this formula
to a particular set of evolution parameters --- one sees that
angular ordering leads to significant reduction of diffusion into the
non-perturbative region. It has been shown by Mueller\cite{Muel} that
diffusion leads to a breakdown in the operator-product expansion
(OPE) at a value of $x$ defined by

\begin{displaymath}
\ln \frac{x_0}{x} \simeq \frac{1}{2 \tchi_c''}\ln \frac{Q^2}{\Lambda^2}
\,,
\end{displaymath}

\noindent where $\Lambda$ is the QCD scale, $x_0$ some starting point
for the small-$x$ evolution, and $Q^2$ the hard scale of the problem.
Accordingly, the effect of angular ordering can be seen as extending
the $x$-range over which the OPE remains valid.

\section{Associated quantities}
The analysis of the gluon distribution is only the first step of a
programme which aims to produce a Monte Carlo event generator based
on the CCFM equation. An intermediate step is to examine associated
final-state properties which are accessible using methods similar to
that used for the structure function. Full details, including a
description of the methods used, will be given in a forthcoming
publication.\cite{bmss:fut} Here, preliminary results  
will be presented for two quantities: the probability of having $n$
emissions with a transverse momentum above a certain minimum; and the
transverse momentum flow as a function of rapidity. In neither case
has the photon-gluon fusion matrix element been included, so the
results are not directly comparable with experiment.

\begin{figure}
\begin{center}
\epsfig{file=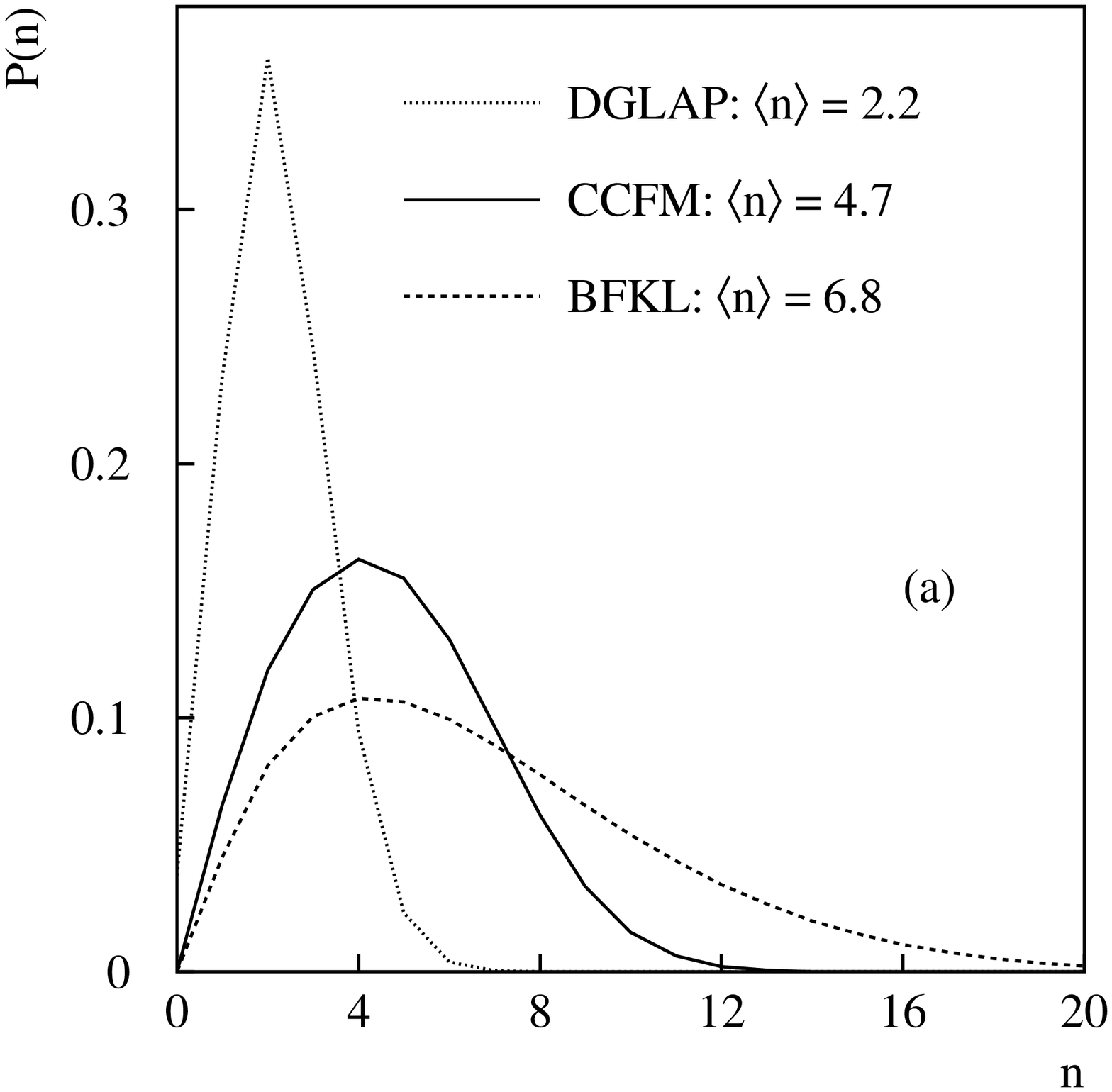,width=0.48\textwidth}
\epsfig{file=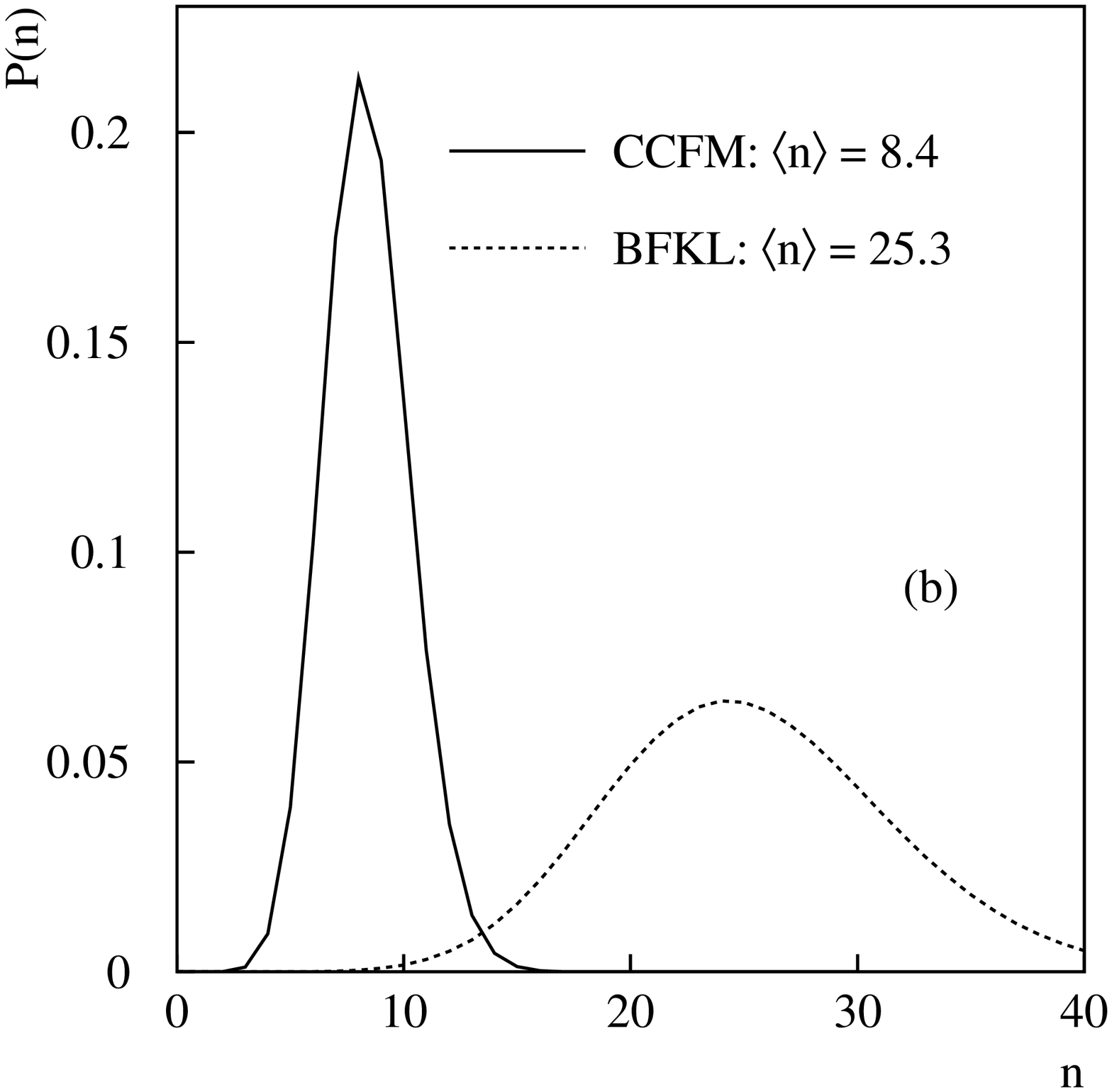,width=0.48\textwidth}
\end{center}
\fcaption{The probability distribution for the number of emissions
with transverse momentum $q>1$~GeV (a) and $q>0.007$~GeV (b); 
$x=5.10^{-5}$, $k=5$~GeV, $\as=0.2$} 
\label{fig:pn}
\end{figure}

The probability of having $n$ primary emitted gluons whose transverse
momentum is larger than some minimum $q>\mu$ is an example of a
quantity which differs between the BFKL and CCFM equations at leading
order:  in the BFKL case, for small $\mu$, it
varies in proportion to $\log \mu$, whereas in the CCFM case, for
small $\mu$ it is independent of $\mu$. Figure~\ref{fig:pn}a shows the
probability distribution for $\mu=1$~GeV, while figure~\ref{fig:pn}a
shows it for $\mu=0.007$~GeV. In the first case the curves are not too
different (e.g. the maxima are at the same point), with the main
difference being that the BFKL case has a 
long tail which is suppressed by phase-space constraints when angular
ordering is introduced. On the other hand, for small $\mu$ the two
distributions are quite different, with their maxima at very different
positions: in the BFKL case there are many small-momentum emissions
which are eliminated by the introduction of angular ordering.

Returning to the phenomenologically more meaningful case of
$\mu=1$~GeV, a simple analytical calculation of the DGLAP result has
been also included:

\begin{equation}
  P(n) = \frac{(\asb \ln k^2 \ln 1/x)^n}{(n!)^2}\,.
\end{equation}

\noindent As one would have expected, the CCFM result lies between the
BFKL and DGLAP cases.

\begin{figure}
\begin{center}
\epsfig{file=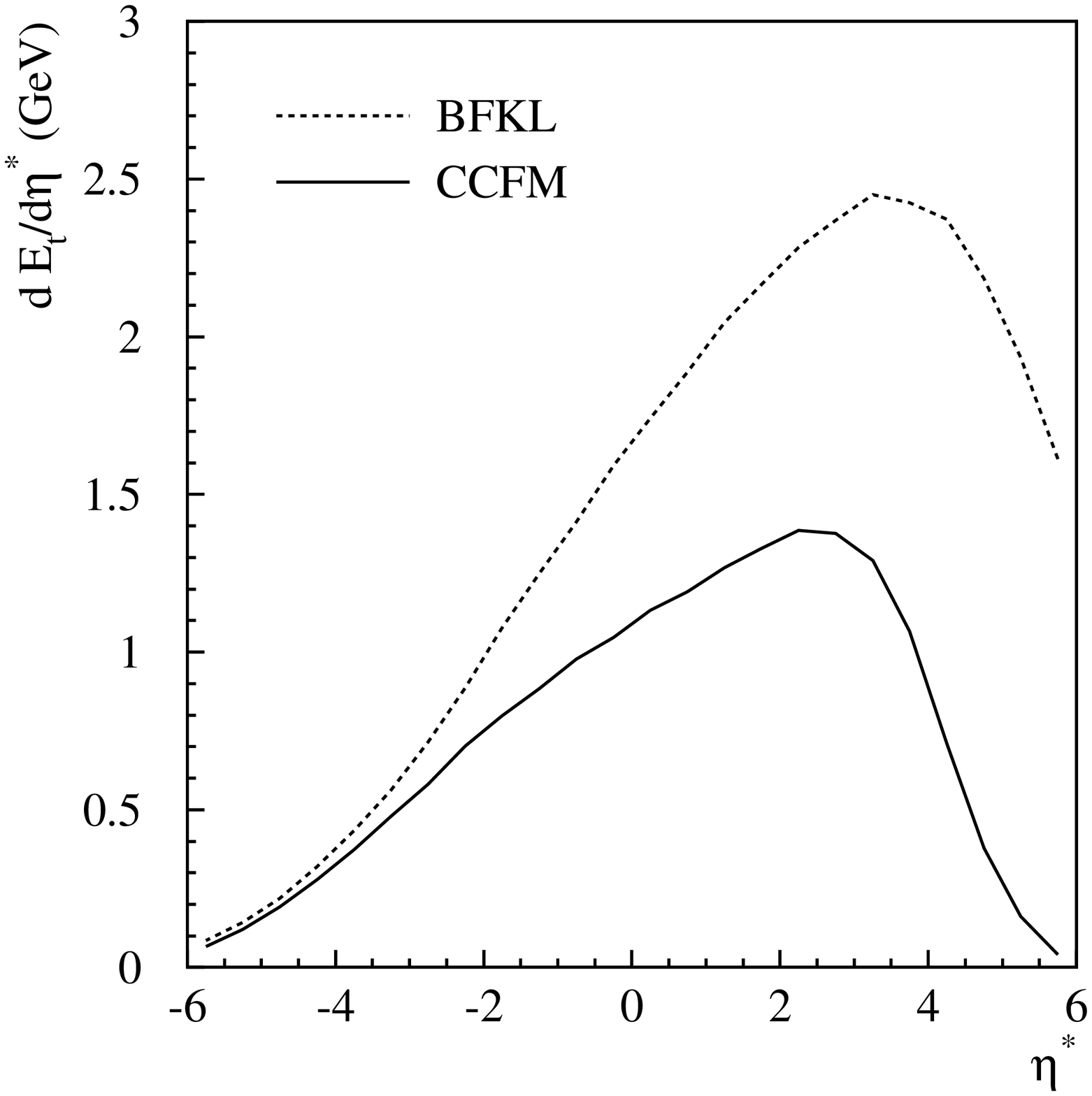,width=0.6\textwidth}
\end{center}
\fcaption{The transverse energy flow in the hadronic centre of mass
frame as a function of the rapidity $\eta^{*}$; the proton direction
is to the left; $x=5.10^{-5}$, $k=5$~GeV, $\as=0.2$.} 
\label{fig:etflow}
\end{figure}

The other quantity for which we have a preliminary result is the
transverse energy flow as a function of rapidity. This is shown in
figure~\ref{fig:etflow} where one sees that angular ordering
significantly reduces the transverse energy flow.

\section{Conclusions and outlook}
In this talk, I have presented results on the effect of angular
ordering on structure functions and two associated
quantities. The main results are to be found in figures~\ref{figcvg}
and \ref{fig:minva}, where it is seen that angular ordering reduces
the height of the minimum of the small-$x$ characteristic function,
shifts its position to larger $\ga$ and strongly reduces its second
derivative (corresponding to a reduction in diffusion).

The technology developed for the study of structure functions is also
being applied to the analysis of associated final-state properties and
some preliminary results have been presented here, illustrating that
angular ordering tends to reduce the number of emissions, particularly
those with low transverse momenta, and that it also reduces the mean
transverse energy flow. 

To carry out phenomenology, certain extra elements are being
implemented, among them the running of $\as$ and the inclusion of
the hard matrix element, which determines the gluon-photon interaction.
Also under development is a backward-evolution CCFM Monte Carlo event
generator. 

\section{Acknowledgements}
This research was carried out in collaboration with G.~Bottazzi,
G.~Marchesini and M.~Scorletti and supported by funding from the
Italian INFN. We would like to thank M. Ciafaloni, Yu.L. Dokshitzer,
A.H. Mueller and B.R. Webber for helpful discussions. I am also
grateful to the Fermilab theory group, where this talk was written up,
for their kind hospitality.

\end{document}